\documentstyle[11pt,newpasp,twoside,epsf]{article}
\begin{document}

\title{Multi-frequency Polarization Imaging of Blazars with Cyclic Activity}

\author{Pyatunina T.B.}
\affil{Insitute of Applied Astronomy of the RAS, Zhdanovskaya St. 8, 197042
St.-Petersburg, Russia}
\author{Gabuzda D.C.}
\affil{Physics Department, University College Cork, Cork, Ireland}
\author{Jorstad S.G.}
\affil{Institute for Astrophysical Research, Boston University, Boston, USA}
\author{Aller M.F. and Aller H.D.}
\affil{Astronomy Department, University of Michigan, USA}
\author{Ter\"asranta H.}
\affil{Mets\"ahovi Radio Observatory, Helsinki University of Technology, Finland}

\begin{abstract}
Results of multi-frequency VLBA polarization imaging for sources
showing evidence of quasi-periodic variations in their total flux density
are discussed. Possible origins of the observed behavior are proposed.
\end{abstract}

Quasi-periodic flux-density variations can be induced,
among other factors, by quasi-periodic perturbations in the
base of the jet (Valtaoja et al. 2000). Such perturbations
are manifest as low-frequency-delayed ``core outbursts'', and lead to
the ejection of new jet components (Pyatunina et al. 2000).
As has been shown by Gabuzda et al. (1994) and Gomez, Marscher, \& Alberdi (1999),
the ejection of new components is sometimes preceeded
by an increase of the core degree of polarization and/or swing of the core
polarization angle $\chi$.
We present preliminary results of observations of some sources that display
quasi-periodic variations of their flux density. Our aim was to investigate
possible connections between these variations and the ejection of new VLBI
components and the role of the magnetic field in various stages of the
components' evolution, and also to determine the total duration of the
activity cycles in these sources.

Note that there is some ambiguity in how to define the duration of the
activity cycles. Physically, this duration can be thought of as the time
interval from the appearance of the primary excitation in the base of the
jet up until the propagating excitation fades and merges with the quiescent
jet emission. However, it may be difficult to determine this time interval
observationally, and in practice it is easier to approximate the duration
of the activity cycle as the interval between successive core outbursts.

The only source for which we have found a reliable connection between the
presence of quasi-periodic flux-density variations and the ejection of new
jet component is 0059+581. The quasi-periodic flares in this source occur
approximately every 2 years, with alternating flares being optically thick
(associated with brightening of the core) and optically thin (associated
with the emergence of new jet components). The extrapolated birth time of the
component that emerged from the core during the 1998--1999 ``thin'' outburst,
$T_0=1996.3$, coincides with the onset of the associated ``thick'' outburst.
Thus, the entire cycle of the quasi-periodic activity lasts $\approx4$~years.
We also predict that a new VLBI component will emerge in position angle
$\simeq 170^{\circ}$ before the end of 2003 (Pyatunina et al. 2003).

Quasi-periodic variations with a time scale of $2.4\pm0.2$~yr have also been
found in the light curves of 0133+476 during 1990--1999.
Analysis of the structural evolution over five years indicates that, as in
the case of 0059+581, these variations are connected with the birth of a
new jet component. However, the variations can be better described by the
precession of the jet nozzle (Pyatunina et al., in prep.) during an early
stage of the activity cycle, before and just after the ejection of a new
VLBI component.

The multi-frequency light curves of 0945+408 show two bright outbursts:
a core outburst near 1986 and a jet outburst near 1994 (Fig.~1). Our
deep 8~GHz VLBA map reveals a faint jet extending up to $\sim25$~mas
\begin{figure}
\plotone{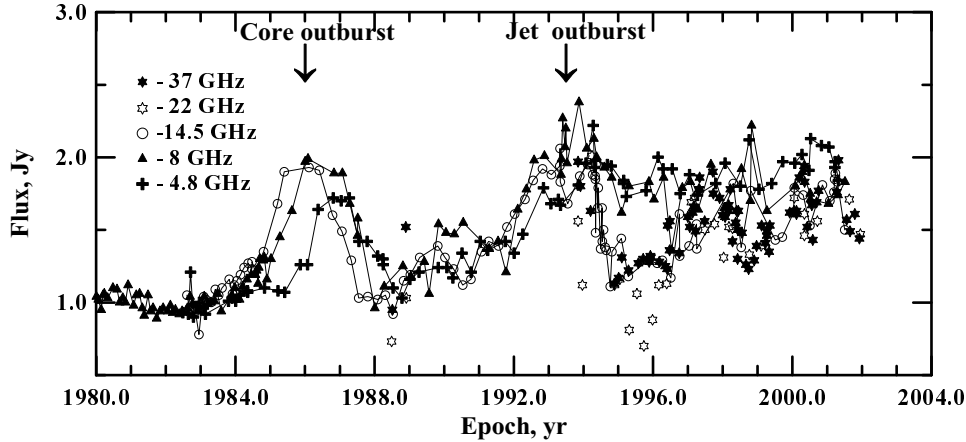}
\caption{Radio light curves of 0945+408.}
\end{figure}
from the core. The structure of the jet resembles a string of beads,
testifiing that the activity in the source can be discrete in time;
i.e., the time interval between the onset of successive activity cycles
can be longer than the duration of the activity itself. The same is probably
true of 0133+476, 1308+326, and 2145+067.
\begin{figure}
\plottwo{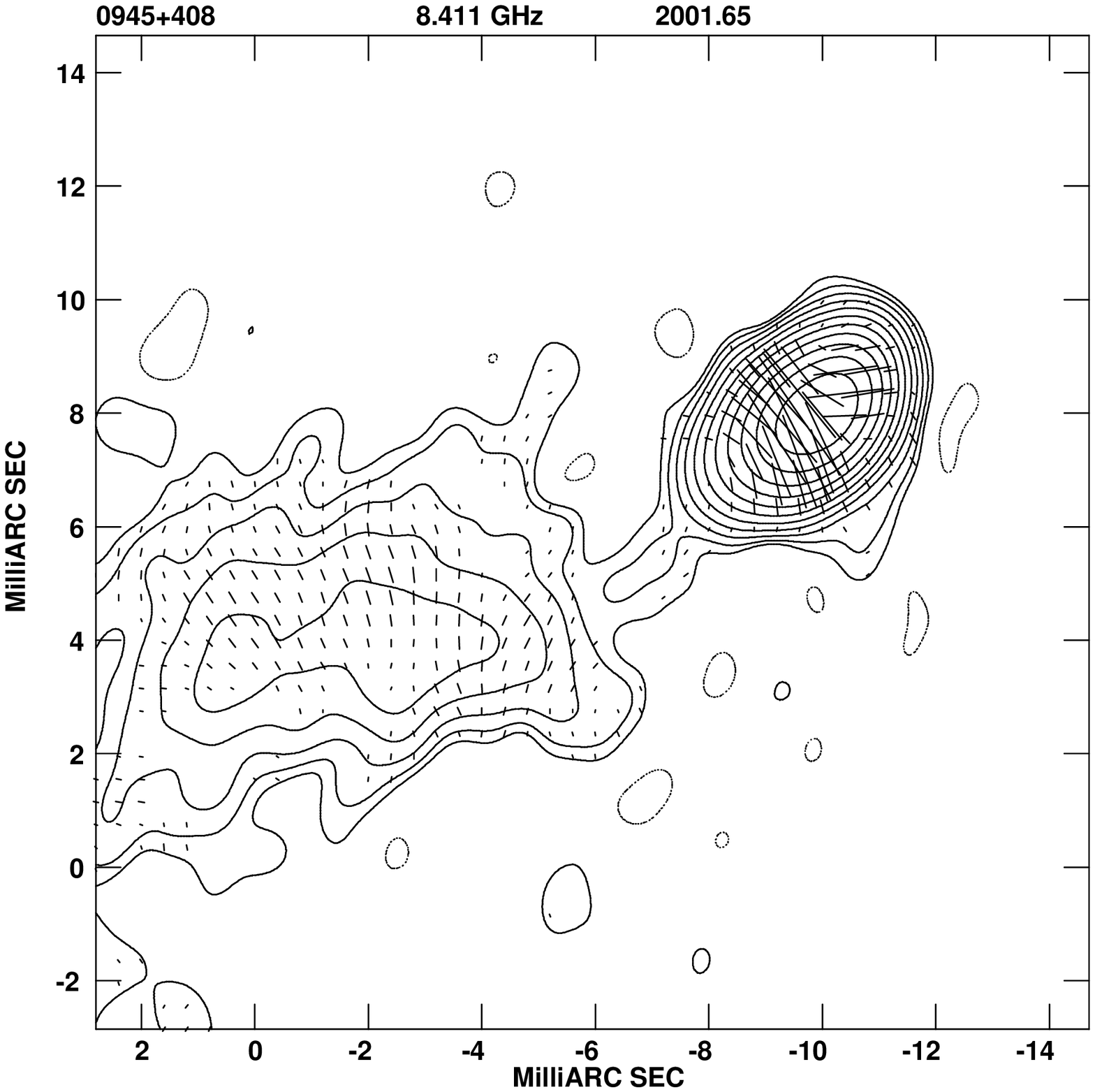}{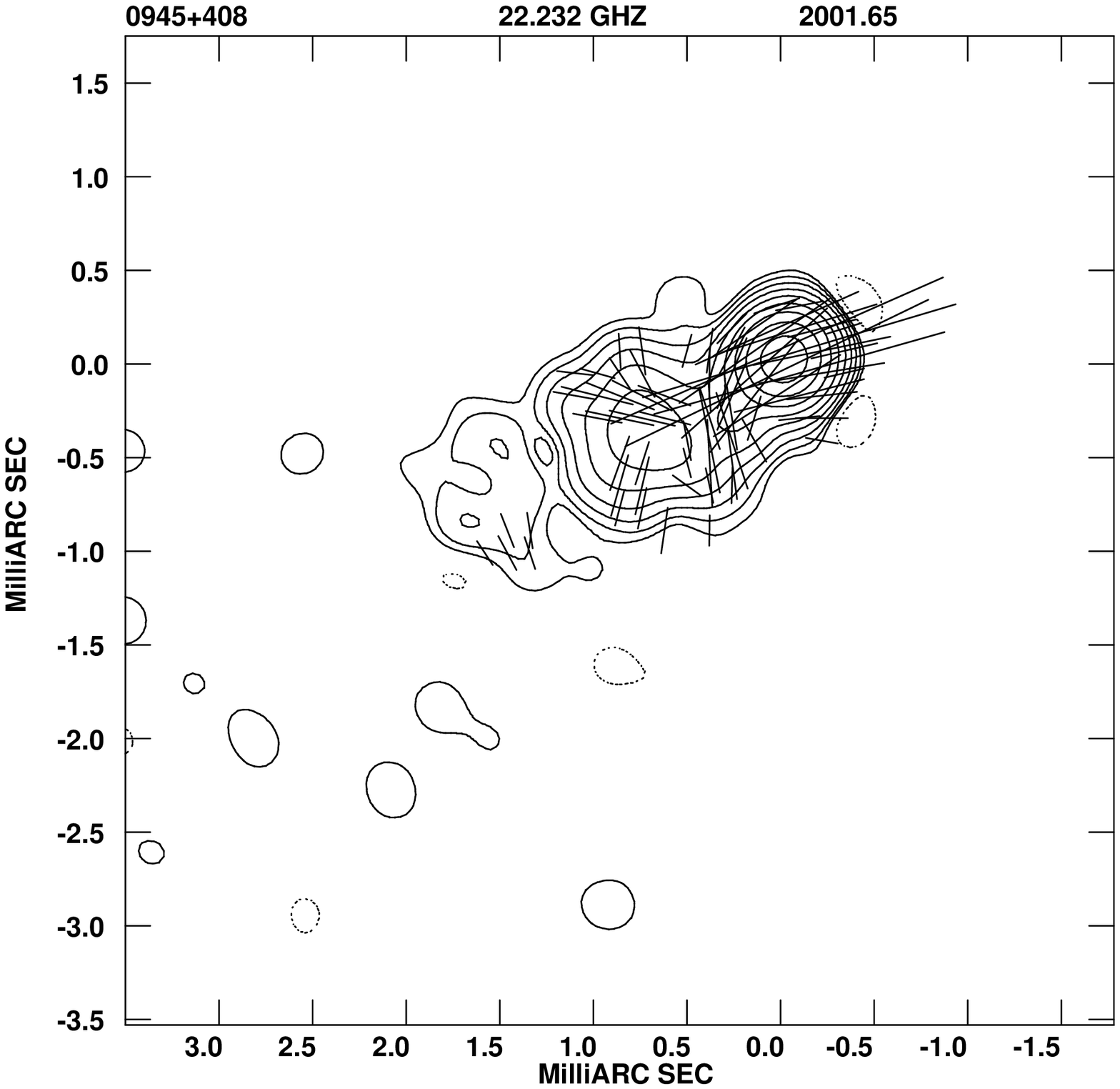}
\caption{8~GHz ({\it left}) and 22~GHz ({\it right}) VLBA maps of 0945+408
at epoch 2001.65. Polarized-intensity vectors are overlaid on total intensity
contours.}
\end{figure}
The higher-resolution maps at 15, 22 and 43~GHz show that the inner jet
of 0945+408 consists of a diffuse outer region and two compact inner
components, indicating that one activity cycle can give birth to several
new jet components. The last two ejections are probably associated with
faint outbursts seen at mm wavelengths near epochs 1997.5 and 2001 (Fig.~1).

The polarization in the extended region of the 8-GHz jet is concentrated
first toward the southern, then toward the northern edge of the visible jet,
and there is a swing in the polarization position angle in the transition
between these two regions.  Assuming the regions of jet emission are optically
thin, the magnetic field is first transverse to the
flow direction, then becomes roughly aligned with the jet flow direction
at the northern edge of the jet. This may suggest interaction with a
surrounding medium; however, a magnetic field structure with a ``spine'' of
transverse field and a ``sheath'' of longitudinal field is also expected
in the case of a helical jet {\bf B} field.  In this latter case, it is
also possible that the magnetic field plays an important role in confining
individual jet components and the jet itself (see Fig. 2).

We have estimated the entire duration of the activity cycles to be
$\ge25$~yr for 0945+408, $\ge20$~yr for 1308+326, and $\ge12$~yr for
0133+476, significantly longer than the $\approx4$~yr period found for
0059+581.

\end{document}